\documentstyle[amssymb]{article}

\begin{document}

\begin{center}
{\large Cuspons, peakons and regular gap solitons between three dispersion
curves}
\end{center}

\bigskip\bigskip

\begin{center}
Roger Grimshaw\footnote{
-mail R.H.J.Grimshaw@lboro.ac.uk}

\smallskip Department of Mathematical Sciences, Loughborough University,

Loughborough, LE11 3TU, UK

\bigskip\smallskip

Boris A. Malomed\footnote{
-mail malomed@eng.tau.ac.il}

\smallskip Department of Interdisciplinary Studies, Faculty of Engineering,

Tel Aviv University, Tel Aviv 69978, Israel

\bigskip

Georg A. Gottwald\footnote{
-mail g.gottwald@surrey.ac.uk}

Department of Mathematics and Statistics, Surrey University, Guildford, GU2
7XH, UK
\end{center}

\newpage

\begin{center}
{\large {\bf ABSTRACT}}
\end{center}

A general model is introduced to describe a wave-envelope system for the
situation when the linear dispersion relation has three branches, which in
the absence of any coupling terms between these branches, would intersect
pair-wise in three nearly-coincident points. The system contains two waves
with a strong linear coupling between them, to which a third wave is then
coupled. This model has two gaps in its linear spectrum. As is typical for
wave-envelope systems, the model also contains a set of cubic nonlinear
terms. Realizations of this model can be made in terms of temporal or
spatial evolution of optical fields in, respectively, either a planar
waveguide, or a bulk-layered medium resembling a photonic-crystal fiber,
which carry a triple spatial Bragg grating. Another physical system
described by the same general model is a set of three internal wave modes in
a density-stratified fluid, whose phase speeds come into close coincidence
for a certain wavenumber. A nonlinear analysis is performed for
zero-velocity solitons, that is, they have zero velocity in the reference
frame in which the third wave has zero group velocity. If one may disregard
the self-phase modulation (SPM) term in the equation for the third wave, we
find an analytical solution which shows that there simultaneously exist two
different families of solitons: regular ones, which may be regarded as a
smooth deformation of the usual gap solitons in a two-wave system, and {\it
cuspons}, which have finite amplitude and energy, but a singularity in the
first derivative at their center. Even in the limit when the linear coupling
of the third wave to the first two nearly vanishes, the soliton family
remains drastically different from that in the uncoupled system; in this
limit, regular solitons whose amplitude exceeds a certain critical value are
replaced by {\it peakons}. While the regular solitons, cuspons, and peakons
are found in an exact analytical form, their stability is tested
numerically, which shows that they all may be stable. If the SPM terms are
retained, we find that there may again simultaneously exist two different
families of generic stable soliton solutions, namely, regular ones and
peakons. Direct simulations show that both types of solitons are stable in
this case.

\bigskip


\newpage

\section{Introduction}

\subsection{The model system}

Gap solitons (GS) is a common name for solitary waves in nonlinear models
which feature one or more \cite{review2} gaps in their linear spectrum, see
review papers \cite{review} and \cite{review2}, respectively, for these two
cases. A soliton may exist if its frequency belongs to the gap, as then it
does not decay into linear waves.

Gaps in the linear spectrum are a generic phenomenon in two- or
multi-component systems, as intersection of dispersion curves belonging to
different components is generically prevented by a linear coupling between
the components. Excluding cases when the system's linear spectrum is
unstable (which are possible in a fluid dynamics application \cite{Ming}),
the intersection avoidance alters the spectrum so that a gap opens in place
of the intersection. Approximating the two dispersion curves, that would
intersect in the absence of the coupling, by straight lines, and assuming a
generic cubic [$\chi ^{(3)}$] nonlinearity, one arrives at a generalized
massive Thirring model (GMTM), which has a family of exact GS solutions that
completely fill the gap \cite{GMTM}. The model has a direct application to
nonlinear optics, describing co-propagation of forward- and
backward-traveling electromagnetic waves in a fiber with a resonant Bragg
grating (BG). Gap solitons, first predicted theoretically, were observed in
experiments with light pulses launched into a short piece of the BG-equipped
fiber \cite{experiment2} (in fact, optical solitons that were first observed
in the BG fiber \cite{experiment} were, strictly speaking, not of the GS
type, but more general ones, whose central frequency did not belong to the
fiber's bandgap).

GS are known not only in optics but also in other physical settings, for
instance, in density-stratified fluid flows, where dispersion curves
pertaining to two different internal-wave modes often exhibit
near-intersections. Again taking into regard weak nonlinearity, one can
predict the occurrence of GS in density-stratified fluids \cite{we}.

In this work, we aim to consider GS that may exist in a generic situation of
the next type, when the underlying system contains three wave components,
and the corresponding dispersion curves intersect at three nearly-coincident
points, unless linear coupling terms are taken into account. Situations of
this type can readily occur in the above-mentioned density-stratified fluid
flows, since tuning of two suitable external parameters can often lead to a
near-coincidence in the linear phase speeds of three independent internal
wave modes, for certain wavenumbers (see \cite{fluid}). Indeed, similar
considerations can be applied to many other physical systems. As for the
usual GS systems, the generic wave-envelope model can be expected to contain
cubic nonlinear terms.

In optics, a $\chi ^{(3)}$-nonlinear model with three linearly coupled waves
is possible too, in terms of either temporary evolution of fields in a
planar nonlinear waveguide equipped with a {\em triple} BG in the form of
three systems of parallel scores, or spatial evolution of stationary fields
in a bulk waveguide with a similar triple BG consisting of three systems of
parallel interfaces between layers. The latter realization seems natural
enough, as it strongly resembles photonic-crystal fibers, which have
recently attracted a great deal of interest \cite{PCF}. Note that the former
version of the model is a generalization of a three-wave model for a $\chi
^{(2)}$-nonlinear planar waveguide with an ordinary BG, which was introduced
in \cite{Mak}. Both versions of the proposed model are illustrated in Fig.
1, where the periodic lattice shows the triple BG. In the case of the
temporal evolution, Fig. 1 displays the planar waveguide, while in the case
of the spatial evolution, it is a transverse cross-section of the bulk
waveguide.

We stress that the lattice in Fig. 1 is not completely symmetric; although
the triangular cells of the lattice are equilateral ones, the two diagonal
sub-gratings are assumed, in the general case, to have the strength 
(contrast of the refractive index)
smaller than the horizontal one. The bold triangles inscribed into the two
triangular cells illustrate the resonant Bragg reflections that give rise to
linear couplings between the waves. Then, neglecting intrinsic dispersion or
diffraction of the waves in comparison with the strong artificial
dispersion/diffraction induced by the Bragg reflections, normalized
equations governing the spatial evolution of the three fields whose Poynting
vectors are shown by three bold arrows in Fig. 1, are 
\begin{equation}
i\left( \frac{\partial u_{1}}{\partial t}-\frac{\partial u_{1}}{\partial x}-
\frac{1}{\sqrt{3}}\frac{\partial u_{1}}{\partial y}\right) +u_{2}+\kappa
u_{3}+\left( |u_{1}|^{2}+2|u_{2}|^{2}+2|u_{3}|^{2}\right) u_{1}=0\,,
\label{01}
\end{equation}
\begin{equation}
i\left( \frac{\partial u_{2}}{\partial t}+\frac{\partial u_{2}}{\partial x}-
\frac{1}{\sqrt{3}}\frac{\partial u_{2}}{\partial y}\right) +u_{1}+\kappa
^{\ast }u_{3}+\left( |u_{2}|^{2}+2|u_{1}|^{2}+2|u_{3}|^{2}\right) u_{2}=0\,,
\label{02}
\end{equation}
\begin{equation}
i\left( \frac{\partial u_{3}}{\partial t}+\frac{2}{\sqrt{3}}\frac{\partial
u_{3}}{\partial y}\right) +\kappa ^{\ast }u_{1}+\kappa u_{2}+\left(
|u_{3}|^{2}+2|u_{1}|^{2}+2|u_{2}|^{2}\right) u_{3}=\omega _{0}u_{3}\,.
\label{03}
\end{equation}
Here, the evolution variable $t$ is the proper time in the case of the
temporal evolution in the planar waveguide, or the coordinate $z$ in the
direction perpendicular to the plane of the figure in the case of the
spatial evolution in the bulk waveguide. In the latter case, the beam enters
the medium through the plane $z=0$ and evolves along the coordinate $z$,
that is why it plays the role of the evolution variable in the equations
(the initial conditions necessary to launch a soliton will be discussed in more
detail below). The relative coefficients in front of the $x$- and $y$-
derivative terms correspond to the geometry in Fig. 1, the coefficient of
the walk-off in the $x$-direction in the first two equations being
normalized to be $\pm 1$. The coefficient of the BG-induced linear
conversion between the waves $u_{1}$ and $u_{2}$ is normalized to be $1$,
while the parameter $\kappa $ (which is complex, in the most general case,
but see the discussion below) accounts for the linear conversion between
these waves and the third wave $u_{3}$, and the usual ratio $1:2$ between
the coefficients of the self-phase modulation (SPM) and cross-phase
modulation (XPM) is adopted. Lastly, $\omega _{0}$ is a frequency/wavenumber
mismatch between the third and the first two waves, which is caused by the
above-mentioned asymmetry between the diagonal and horizontal subgratings,
as well as by other reasons.

As we mentioned above, the Bragg constant $\kappa $ in Eqs. (\ref{01}) - 
(\ref{03}), which couples the field $u_{3}$ to the pair $u_{1,2}$, is complex
in the general case (note that the constant of the Bragg coupling between
the fields $u_{1}$ and $u_{2}$ might also be complex in its primary form,
and making it equal to $1$ in Eqs. (\ref{01}) and (\ref{02}) involves
opposite constant phase shifts of the fields $u_{1}$ and $u_{2}$, which is
why $\kappa $ and $\kappa ^{\ast }$ appear exactly as in Eqs. (\ref{01}) - 
(\ref{03})). However, assuming that each score, the families of which
constitute the triple grating (in the case of the temporal-domain evolution
in the planar waveguide), gives rise to simple reflection described by the
classical Fresnel formulas, it is easy to conclude that all the coupling
constants are real and positive, provided that either the light is polarized
orthogonally to the waveguide's plane, and the reflection takes place from a
less optically dense material (i.e., the ``score'' is, literally, a shallow
trough on the surface of the planar waveguide), or the light is polarized
parallel to the plane of the waveguide, and the reflection is from a more
optically dense material. Similarly, in the case when the same equations
describe the spatial evolution of the optical fields in the\ layered bulk
medium, one may assume that either the light is polarized in the $z$
-direction, and seams between the layers are filled with a material (for
instance, air) which is optically less dense than the bulk medium, or the
polarization is orthogonal to the $z$-axis (i.e., it is parallel to the
plane of Fig. 1), and the material filling the inter-layer seams is
optically denser than the host medium.

In the present paper, we focus on this case, which was described above in
detail for the realizations of the model in terms of both planar and bulk
optical waveguides, and which corresponds to $\kappa $ real and positive in
Eqs. (\ref{01}) - (\ref{03}). Note, incidentally, that the case when $\kappa 
$ is real and {\em negative} can be reduced to the same case simply by
reversing the sign in the definition of $u_{3}$.

The model displayed in Fig. 1 may be further generalized by introducing an
additional asymmetry, which will remove the equality between the horizontal
side of the lattice's triangular cell and its diagonal sides. Then, the
simultaneous fulfillment of the Bragg-reflection conditions for the waves 
$u_{1,2}$ and $u_{3}$ can be secured by making the waveguide anisotropic.
However, such a generalization goes beyond the scope of this paper.

For the physical realization of the model, Eqs. (\ref{01}) - (\ref{03}) must
be supplemented by initial conditions at $t=0$ in the case of the temporal
evolution in the planar waveguide, or boundary conditions at $z=0$ in the
case of the spatial evolution in the bulk medium. It is sufficient to assume
that, at $t=0$, a single wave component (for instance, $u_{3}$) is launched
into the waveguide. The linear-coupling terms in the equations will then
start to generate the other components, and, if solitons that might exist in
this model are stable (see below), they may self-trap from such an initial
beam.

Bearing in mind also the above-mentioned application to internal waves in
stratified fluids, as well as similar realizations in other physical media,
Eqs. (\ref{01}) - (\ref{03}) may be naturally extended by introducing more
general SPM and XPM coefficients, as in applications other than nonlinear
optics, the ratios between the XPM and SPM\ coefficients may be different
from those adopted above. Thus the generalized system of equations takes the
following form, in which we confine consideration to $y$-independent
solutions (the consideration of possible three-dimensional solitons in the
case of $y$-dependent fields is not an objective of this work), 
\begin{equation}
i(\frac{\partial u_{1}}{\partial t}-\frac{\partial u_{1}}{\partial x}
)+u_{2}+\kappa u_{3}+\alpha \left( \alpha \sigma
_{1}|u_{1}|^{2}+\alpha|u_{2}|^{2}+|u_{3}|^{2}\right) u_{1}=0\,,  \label{1}
\end{equation}
\begin{equation}
i(\frac{\partial u_{2}}{\partial t}+\frac{\partial u_{2}}{\partial x}
)+u_{1}+\kappa u_{3}+\alpha \left( \alpha \sigma
_{1}|u_{2}|^{2}+\alpha|u_{1}|^{2}+|u_{3}|^{2}\right) u_{2}=0\,,  \label{2}
\end{equation}
\begin{equation}
i\frac{\partial u_{3}}{\partial t}+\kappa \left( u_{1}+u_{2}\right)
+\left(\sigma _{3}|u_{3}|^{2}+\alpha |u_{1}|^{2}+\alpha
|u_{2}|^{2}\right)u_{3}=\omega _{0}u_{3}\,.  \label{3}
\end{equation}
where, in accord with the discussion above, we set $\kappa $ to be real and
positive.

The coefficients $\sigma _{1,3}$ and $\alpha $ in Eqs. (\ref{1}) - (\ref{3})
are the generalized SPM and XPM coefficients, respectively. In particular, $
\alpha $ is defined as a relative XPM coefficient between the first two
waves and the third wave. In fact, the coefficients $\sigma _{1}$ and $
\sigma _{3}$ both may be normalized to be $\pm 1$, unless they are equal to
zero; however, it will be convenient to keep them as free parameters, see
below (note that the SPM coefficients are always positive in the optical
models, but in those describing density-stratified fluids they may have
either sign). In optical models, all the coefficients $\alpha $ and $\sigma
_{1,3}$ are positive. However, in the models describing the internal waves
in stratified fluids, there is no inherent restriction on their signs, and
some of them may indeed be negative.

The symmetry between the walk-off terms in Eqs. (\ref{1}) and (\ref{2}) is
not really essential, and we will comment later on the more general case
when these terms are generalized as follows: 
\begin{equation}
-\frac{\partial u_{1}}{\partial x}\,\rightarrow -c_{1}\frac{\partial u_{1}}
{\partial x}\,,\,+\frac{\partial u_{2}}{\partial x}\,\rightarrow +c_{2}\frac{
\partial u_{1}}{\partial x}\,,  \label{asymmetry}
\end{equation}
where $c_{1}$ and $c_{2}$ are different, but have the same sign. As for Eq.
(ref{3}), it is obvious that the walk-off term in this equation, if any, can
always be eliminated by means of a straightforward transformation.

We have kept only the most natural nonlinear SPM and XPM terms in Eqs. (\ref
{1}) - (\ref{3}), i.e., the terms of the same types as in the standard GMT
model. Additional terms, including nonlinear corrections to the linear
couplings (e.g., a term $\sim |u_{1}|^{2}u_{2}$ in Eq. (\ref{1})) may appear
in more general models, such as a model of a deep (strong) BG \cite{deep}.

Equations (\ref{1}) - (\ref{3}) conserve the norm, which has the physical
meaning of energy in optics, 
\begin{equation}
N\,\equiv \,\sum_{n=1,2,3}\int_{-\infty }^{+\infty }\left|
u_{n}(x)\right|^{2}dx,  \label{E}
\end{equation}
the Hamiltonian, 
\begin{eqnarray}  \label{focus}
H &\equiv &H_{{\rm grad}}+H_{{\rm coupl}}\,+H_{{\rm focus}},  \label{H} \\
H_{{\rm grad}} &\equiv &\frac{i}{2}\int_{-\infty }^{+\infty
}\left(u_{1}^{\ast }\frac{\partial u_{1}}{\partial x}-u_{2}^{\ast }\frac{
\partial u_{2}}{\partial x}\right) dx+{\rm c.c.},  \label{grad} \\
H_{{\rm coupl}} &\equiv &-\int_{-\infty }^{+\infty }\left[
u_{1}^{\ast}u_{2}+\kappa u_{3}^{\ast }\left( u_{1}+u_{2}\right) \right] dx+
{\rm c.c.},  \label{coupl} \\
H_{{\rm focus}} &\equiv &-\int_{-\infty }^{+\infty }\left[ \frac{1}{2}
\alpha^{2}\sigma _{1}\left( \left| u_{1}\right| ^{4}+\left|
u_{2}\right|^{4}\right) +\frac{1}{2}\sigma _{3}\left| u_{3}\right|
^{4}+\alpha^{2}\left| u_{1}\right| ^{2}\left| u_{2}\right| ^{2}+\alpha
\left|u_{3}\right| ^{2}\left( \left| u_{1}\right| ^{2}+\left|
u_{2}\right|^{2}\right) \right] dx\,,
\end{eqnarray}
and the momentum, which will not be used here. In these expressions, the
asterisk and ${\rm c.c.}$ both stand for complex conjugation, $H_{{\rm grad}
} $, $H_{{\rm coupl}}$, and $H_{{\rm focus}}$ being the gradient,
linear-coupling, and self-focusing parts of the Hamiltonian. To obtain the
Eqs. (\ref{1}) - (\ref{3})from the Hamiltonian, the conjugate pairs of the
variables are defined, in a standard fashion, as $u_n,u_n^{*}$.

Our objective is to find various types of solitons existing in the generic
three-wave system (\ref{1}) - (\ref{3}) and investigate their stability.
Focusing first on the case (suggested by the analogy with GMTM) when the SPM
term in Eq. (\ref{3}) may be neglected (i.e., $\sigma _{3}=0$), in section 3
we find a general family of zero-velocity solitons in an exact analytical
form. We will demonstrate that they are of two drastically different types:
regular GS, and {\it cuspons}, i.e., solitons with a cusp singularity at the
center, in which the soliton amplitude is finite, but the derivative is
infinite; further, the energy of the cuspons is finite. Cuspons are known to
exist in degenerate models without linear terms (except for the evolution
term such as $\partial u/\partial t$), i.e., without a linear spectrum, a
well-known example being the exactly integrable Camassa-Holm (CH) equation 
\cite{CH,cuspon} (see also \cite{Fokas}). Our model resembles the CH one in
the sense that both give rise to coexisting solutions in the form of regular
solitons and cuspons. The cause for the existence of these singular solitons
in our model is the fact that, looking for a zero-velocity soliton solution,
one may eliminate the field $u_{3}$ by means of an algebraic relation
following, in this case, from Eq. (\ref{3}). The subsequent substitution of
that result into the first two equations (\ref{1}) and (\ref{2}) produces a 
{\it rational} nonlinearity in them, the corresponding rational functions
featuring a singularity at some (critical) value of the soliton's amplitude.
If the amplitude of a regular-soliton solution is going to exceed the
critical value, it actually cannot exist, and in the case when $\sigma
_{3}=0 $ it is replaced by a cuspon, whose amplitude is exactly equal to the
critical value.

In the limit $\kappa \rightarrow 0$, which corresponds to the vanishing
linear coupling between the first two and third waves, the cuspon resembles
a {\it peakon}, which is a finite-amplitude solitary wave with a jump of its
first derivative at the center. Note that peakon solutions, coexisting with
regular solitons (this property is shared by our model), are known in a
slightly different (also integrable) version of the CH equation, see, e.g.,
Refs. \cite{CH,peakon,additional}. We also note that soliton solutions with
a discontinuity in the first derivative have been found in the BG model
(which does contain a linear part) in the case where the grating parameter
changes abruptly \cite{Broderick}.

Then, we show that, when the SPM term is restored in Eq. (\ref{3}) (i.e., 
$\sigma _{3}\neq 0$; the presence or absence of the SPM terms $\sim \sigma
_{1}$ in Eqs. (\ref{1}) and (\ref{2}) is not crucially important), the
system supports a different set of soliton solutions. These are regular GS
and, depending on the sign of certain parameters, a family of peakons,
which, this time, appear as generic solutions, unlike the case $\sigma
_{3}=0 $, when they only exist as a limiting form of the cuspon solutions
corresponding to $\kappa \rightarrow 0$. As far as we know, the model
formulated in the present work is the first spatially uniform non-degenerate
one (i.e., a model with a non-vanishing linear part) which yields both
cuspons and peakons.

\subsection{Stability of the solitons and spatiotemporal collapse}

As concerns the dynamical stability of the various solitons in the model 
(\ref{1}) -(\ref{3}), in this work we limit ourselves to 
direct simulations, as a
more rigorous approach, based on numerical analysis of the corresponding
linear stability-eigenvalue problem \cite{accurate}, is technically
difficult in the case of cuspons and peakons (results of such an analysis,
based on the Evans-function technique, will be presented elsewhere). In
fact, direct simulations of perturbed cuspons and peakons is a hard problem
too, but we have concluded that identical results concerning the stability
are produced (see section 3 below) by high-accuracy finite-difference and
pseudo-spectral methods (each being implemented in more than one particular
form), which lends the results credibility. A general conclusion is that the
regular solitons are always stable. As for the cuspons and peakons, they may
be either stable or unstable.

If the cusp is strong enough, the numerical results presented below
demonstrate that the instability of a cuspon initiates formation of a
genuine singularity, i.e., onset of a {\it spatiotemporal collapse} \cite
{collapse} in the present one-dimensional model. Before proceeding to the
consideration of solitons in the following sections, it is relevant to
discuss collapse phenomenon in some detail.

A simple virial-type estimate for the possibility of the collapse can be
done, assuming that the field focuses itself in a narrow spot with a size 
$(t)$, amplitude $\aleph (t)$, and a characteristic value $K(t)$ of the
field's wavenumber \cite{collapse}. The conservation of the norm (\ref{E})
imposes a restriction $\aleph ^{2}L\sim N$, i.e., $L\sim N/\aleph ^{2}$.
Next, the self-focusing part (\ref{focus}) of the Hamiltonian (\ref{H}),
which drives the collapse, can be estimated as 
\begin{equation}
H_{{\rm focus}}\sim -\aleph ^{4}L\sim -N\aleph ^{2}.  \label{collapsing}
\end{equation}
On the other hand, the collapse can be checked by the gradient term (\ref
{grad}) in the full Hamiltonian, that, in the same approximation, can be
estimated as $H_{{\rm grad}}\sim \aleph ^{2}KL\sim NK$. Further, Eqs. (\ref
{1}) - (\ref{3}) suggest an estimate $K\sim \aleph ^{2}$ for a
characteristic wavenumber of the wave field [the same estimate for $K$
follows from an expression (\ref{phi}) for the exact stationary-soliton
solution given below], thus we have $H_{{\rm grad}}\sim N\aleph ^{2}$.
Comparing this with the expression (\ref{collapsing}), one concludes that
the parts of the Hamiltonian promoting and inhibiting the collapse scale the
same way as $\aleph \rightarrow \infty $ (or $L\rightarrow 0$), hence a {\it
weak collapse} \cite{collapse} may be possible (but does not necessarily
take place) in systems of the present type. We stress that, in
one-dimensional models of GS studied thus far and based on GMTM, collapse
has never been reported. The {\em real existence} of the collapse in the
present one-dimensional three-wave GS model, which will be shown in detail
below as a result of numerical simulations, is therefore a novel dynamical
feature, and it seems quite natural that cuspons and peakons, in the case
when they are unstable, play the role of catalysts stimulating the onset of
the collapse. The possibility of a real collapse in a 1D system is quite
interesting by itself, and also because experimental observation of
spatio-temporal self-focusing in nonlinear optical media is a subject of
considerable interest, see, e.g., Ref. \cite{Bar-Ad}.

\section{Analytical solutions}

\subsection{The dispersion relation}

The first step in the investigation of the system is to understand its
linear spectrum. Substituting $u_{1,2,3}\sim \exp (ikx-i\omega t)$ into Eqs.
(\ref{1} -\ref{3}), and omitting nonlinear terms, we arrive at a dispersion
equation, 
\begin{equation}
(\omega ^{2}-k^{2}-1)(\omega -\omega _{0})\,=\,2\kappa ^{2}(\omega -1).
\label{dispersion}
\end{equation}
If $\kappa =0$, the third wave decouples, and the coupling between the first
two waves produces a commonly known gap, so that the solutions\ to Eq. (\ref
{dispersion}) are $\omega _{1,2}=\pm \sqrt{1+k^{2}}$ and $\omega _{3}=\omega
_{0}$. If $\kappa \neq 0$, the spectrum can be easily understood by treating 
$\kappa $ as a small parameter. However, the following analysis is valid for
all values of $\kappa $ in the range $0<\kappa ^{2}<1$.

First, consider the situation when $k=0$. Three solutions of Eq. (\ref
{dispersion}) are then 
\begin{equation}
\omega =1,\,\omega \,=\,\omega _{\pm }\equiv \,(\omega _{0}-1)/2\pm \sqrt{
(\omega _{0}+1)^{2}/4+2\kappa ^{2}}.  \label{gap0}
\end{equation}
It can be easily shown that $\omega _{-}<\min \{\omega _{0},-1\}\leq \max
\{\omega _{0},-1\}<\omega _{+}$, so that one always has $\omega _{-}<-1$,
while $\omega _{+} < 1$ if $\omega _{0}< 1-\kappa ^{2}$, and
$\omega _{+} > 1$ if $\omega _{0}> 1-\kappa ^{2}$. Next,
it is readily seen that, as $k^{2}\rightarrow \infty $, either $\omega
^{2}\approx k^{2}$, or $\omega \approx \omega _{0}$. Each branch of the
dispersion relation generated by Eq. (\ref{dispersion}) is a monotonic
function of $k^{2}$. Generic examples of the spectrum are shown in Fig. 2,
where the panels (a) and (b) pertain, respectively, to the cases $\omega
_{0}<1-\kappa ^{2}$ with $\omega _{+}<1$, and $\omega _{0}>1$ with $\omega
_{+}>1$. The intermediate case, $1-\kappa ^{2}<\omega _{0}<1$, is similar to
that shown in panel (a), but with the points $\omega _{+}$ and $1$ at $k=0$
interchanged. When $\omega _{0}<1$, the upper gap in the spectrum is $\min
\{\omega _{+},1\}<\omega <$ $\max \{\omega _{+},1\}$, while the lower gap is 
$\omega _{-}<\omega <\omega _{0}$. When $\omega _{0}>1$, the upper gap is 
$\omega _{0}<\omega <\omega _{+}$, and the lower one is $\omega _{-}<\omega
<1 $.

\subsection{Gap solitons}

The next step is to search for GS solutions to the full nonlinear system. In
this work, we confine ourselves to the case of zero-velocity GS,
substituting into Eqs. (\ref{1}) - (\ref{3}) 
\begin{equation}
u_{n}(x,t)\,=\,U_{n}(x)\exp (-i\omega t)\,,\,n=1,2,3,  \label{stationary}
\end{equation}
where it is assumed that the soliton's frequency $\omega $ belongs to one of
the gaps. In fact, even the description of zero-velocity solitons is quite
complicated. Note, however, that if one sets $\kappa =0$ in Eqs. (\ref{1}) -
(\ref{3}), keeping nonlinear XPM couplings between the first two and third
waves, the gap which exists in the two-wave GMT model remains unchanged, and
the corresponding family of GS solutions does not essentially alter, in
accord with the principle that nonlinear couplings cannot alter gaps or open
a new one if the linear coupling is absent \cite{we2}; nevertheless, the
situation is essentially different if $\kappa $ is vanishingly small, but
not exactly equal to zero, see below.

The substitution of (\ref{stationary}) into Eqs. (\ref{1}) and (\ref{2})
leads to a system of two ordinary differential equations for $U_{1}(x)$ and 
$U_{2}(x)$, and an algebraic relation for $U_{3}(x)$, 
\begin{equation}
iU_{1}^{\prime }\,=\omega U_{1}+U_{2}+\kappa U_{3}+\alpha \left(
\alpha\sigma _{1}|U_{1}|^{2}+\alpha |U_{2}|^{2}+|U_{3}|^{2}\right) U_{1},
\label{U1}
\end{equation}
\begin{equation}
-iU_{2}^{\prime }\,=\omega U_{2}+U_{1}+\kappa U_{3}+\alpha \left(
\alpha\sigma _{1}|U_{2}|^{2}+\alpha |U_{1}|^{2}+|U_{3}|^{2}\right) U_{2},
\label{U2}
\end{equation}
\begin{equation}
(\omega _{0}-\omega +\sigma _{3}|U_{3}|^{2}+\alpha
|U_{1}|^{2}+\alpha|U_{2}|^{2})U_{3}\,=\,\kappa (U_{1}+U_{2}),  \label{U3}
\end{equation}
where the prime stands for $d/dx$. To solve these equations, we substitute 
$U_{1,2}=A_{1,2}(x)\exp \left( i\phi _{1,2}(x)\right) $ with real $A_{n}$ and 
$\phi _{n}$. After substituting the expression (\ref{U3}) into equations 
(\ref{U1},\ref{U2}), and some simple manipulations, it can be found that 
$\left( A_{1}^{2}-A_{2}^{2}\right) ^{\prime }=0$ and $\left( \phi _{1}+\phi
_{2}\right) ^{\prime }=0$. Using the condition that the soliton fields
vanish at infinity, we immediately conclude that 
\begin{equation}
A_{1}^{2}(x)=A_{2}^{2}(x)\equiv S(x);  \label{A1^2A2^2}
\end{equation}
as for the constant value of $\phi _{1}+\phi _{2}$, it may be set equal to
zero without loss of generality, so that $\phi _{1}(x)=-\phi _{2}(x)\equiv
\phi (x)/2$, where $\phi (x)$ is the relative phase of the two fields. After
this, we obtain two equations for $S(x)$ and $\phi (x)$ from Eqs. (\ref{U1})
and (\ref{U2}), 
\begin{equation}
\phi ^{\prime }=-2\omega -2\cos \phi -2\alpha ^{2}\left( 1+\sigma_{1}\right)
S-S^{-1}U_{3}^{2}\left( \omega _{0}-\omega -\sigma_{3}U_{3}^{2}\right) \,,
\label{phi}
\end{equation}
\begin{equation}
S^{\prime }=-2S\sin \phi -2\kappa \sqrt{S}U_{3}\sin \left( \phi /2\right) ,
\label{S}
\end{equation}
and Eq. (\ref{U3}) for the third wave $U_{3}$ takes the form of a cubic
algebraic equation 
\begin{equation}
U_{3}\left( \omega _{0}-\omega -2\alpha S-\sigma
_{3}|U_{3}|^{2}\right)\,=\,2\kappa \sqrt{S}\cos \left( \phi /2\right) ,
\label{U3real}
\end{equation}
from which it follows that $U_{3}$ is a real-valued function.

This analytical consideration can be readily extended for more general
equations (\ref{1}) and (\ref{2}) that do not assume the symmetry between
the waves $u_{1}$ and $u_{2}$, i.e., with the group-velocity terms in the
equations altered as in Eq. (\ref{asymmetry}). In particular, the relation 
(\ref{A1^2A2^2}) is then replaced by $c_{1}A_{1}^{2}(x)=c_{2}A_{2}^{2}(x)
\equiv S(x)$. The subsequent analysis is similar to that above, and leads to
results for the asymmetric model that are qualitatively similar to those
presented below for the symmetric case.

Equations (\ref{phi}) and (\ref{S}) have a Hamiltonian structure, as they
can be represented in the form 
\begin{equation}
\frac{dS}{dx}\,=\,\frac{\partial H}{\partial \phi }\,,\quad \frac{d\phi }{dx}
\,=\,-\frac{\partial H}{\partial S}\,,  \label{canonical}
\end{equation}
with the Hamiltonian 
\begin{equation}
H\,=\,2S\cos \phi +\alpha ^{2}\left( 1+\sigma _{1}\right) S^{2}+2\omega
S+U_{3}^{2}\left( \omega _{0}-\omega -2\alpha S\right) -\frac{3}{2}\sigma
_{3}U_{3}^{4}\,,  \label{Hamiltonian}
\end{equation}
which is precisely a reduction of the Hamiltonian (\ref{H}) of the original
system (\ref{1}) - (\ref{3}) for the solutions of the present type. Note
that $H$ is here regarded as a function of $S$ and $\phi $, and the relation
(\ref{U3real}) is regarded as determining $U_{3}$ in terms of $S$ and $\phi$. 
We stress that the dependence $U_{3}(S,\phi )$ was taken into account when
deriving the Hamiltonian representation (\ref{canonical}).

For soliton solutions, the boundary conditions at $x=\pm \infty $ yield $H=0$
so that the solutions can be obtained in an implicit form, 
\begin{equation}
2S\cos \phi +\alpha ^{2}\left( 1+\sigma _{1}\right) S^{2}+2\omega
S+U_{3}^{2}\left( \omega _{0}-\omega -2\alpha S\right) -\left( 3/2\right)
\sigma _{3}U_{3}^{4}\,=\,0.  \label{integral}
\end{equation}
In principle, one can use the relations (\ref{U3real}) and (\ref{integral})
to eliminate $U_{3}$ and $\phi $ and so obtain a single equation for $S$.
However, this is not easily done unless $\sigma _{3}=0$ (no SPM term in Eq.
(ref{3})), and so we proceed to examine this special, but important, case
first. Note that the no-SPM case also plays an important role for GMTM,
which is exactly integrable by means of the inverse scattering transform
just in this case \cite{review}.

\subsection{Cuspons, the case $\protect\sigma_3 = 0$}

Setting $\sigma _{3}=0$ makes it possible to solve Eq. (\ref{U3real}) for 
$_{3}$ explicitly in terms of $S$ and $\phi $, 
\begin{equation}
U_{3}\,=\,\frac{2\kappa \sqrt{S}\cos \left( \phi /2\right) }{
\omega_{0}-\omega -2\alpha S}\,.  \label{explicit}
\end{equation}
For simplicity, we also set $\sigma _{1}=0$ in Eqs. (\ref{1}) and (\ref{2})
and subsequent equations, although the latter assumption is not crucially
important for the analysis developed below. Indeed, the analysis is based on
the fact that the field $U_{3}$ can be explicitly eliminated by means of Eq.
(\ref{explicit}), which is not affected by $\sigma _{1}$. If $\sigma _{1}$
is kept in the system, it merely renormalizes some coefficients in the
formulas derived below.

At the next step, one can also eliminate $\phi $, using Eqs. (\ref{integral}
) and (\ref{explicit}), to derive a single equation for $S$, 
\begin{equation}
\left( dS/dx\right) ^{2}\,=\,4S^{2}F(S),  \label{S1}
\end{equation}
\begin{equation}
F(S)\equiv (1-\omega -\frac{1}{2}\alpha ^{2}S)\left[ 2\left( 1+\frac{
\kappa^{2}}{\omega _{0}-\omega -2\alpha S}\right) \,-\,(1-\omega -\frac{1}{2}
\alpha ^{2}S)\right] .  \label{F}
\end{equation}
The function $F(S)$ has either one or three real zeros $S_{0}$. One is 
\begin{equation}
S_{01}\,=\,2\,\left( 1-\omega \right) /\alpha ^{2},  \label{S01}
\end{equation}
and the remaining two, if they exist, are real roots of the quadratic
equation, 
\begin{equation}
(2+2\omega +\alpha ^{2}S_{0})(\omega _{0}-\omega -2\alpha
S_{0})+4\kappa^{2}\,=0.  \label{quadr}
\end{equation}
Only the smallest positive real root of Eq. (\ref{quadr}), to be denoted 
$S_{02}$ (if such exists), will be relevant below. Note, incidentally, that 
$F(S)$ cannot have double roots. It is easy to see that a consequence of this
fact is that Eq. (\ref{S1}) cannot generate kink solutions, which have
different limits as $x\rightarrow \pm \infty $, for both of which the
right-hand-side of (\ref{S1}) must have a double zero.

For a bright-soliton solution of (\ref{S1}), we need first that $F(0)>0$ (in
this work, we do not consider dark solitons, nor ``anti-dark'' solitons,
i.e., solitons on top of a finite-amplitude flat background, a reason being
that there is little chance that the flat background would be modulationally
stable). Comparing the condition $F(0)>0$ with the expressions given in
Section 2.1 for the gaps in the linear spectrum, it is readily shown that
this condition is exactly equivalent to requiring that $\omega $ belongs to
either the upper or the lower gap of the linear spectrum. We note that the
coupling to the third wave gives rise to nonlinearity of the rational type
in the expression (\ref{F}), despite the fact that the underlying system 
(\ref{1}) - (\ref{3}) contains only cubic polynomial nonlinear terms. Even if
the coupling constant $\kappa $ is small, it is clear that the rational
nonlinearity may produce a strong effect in a vicinity of a {\it critical
value} of the squared amplitude at which the denominator in the expression 
(\ref{F}) vanishes, 
\begin{equation}
S_{{\rm cr}}\,=\left( \,\omega _{0}-\omega \right) /2\alpha ,
\label{critical}
\end{equation}
where one must have $\alpha (\omega _{0}-\omega )>0$ (otherwise, this
critical value is not relevant).

If $S_{{\rm cr}} > 0$, the structure of the soliton crucially depends on
whether, with an increase of $S$, the function $F(S)$ defined by Eq. (\ref{F}
) first reaches zero at $S=S_{0} >0$ (i.e. either $S = S_{01}$ or $S =
S_{02} $, whichever is the smaller positive value), or, instead, it first
reaches the singularity at $S=S_{{\rm cr}}$, i.e., whether $0<S_{0}<S_{{\rm
cr}}$, or $0<S_{{\rm cr}}<S_{0}$. In the former case, the existence of 
$S_{{\rm cr}} $ plays no role, and the soliton is a regular one, having the
amplitude $\sqrt{S_{0}}$. This soliton may be regarded as obtained by a
smooth deformation from the usual GS known in GMTM at $\kappa =0$.

In the case $0<S_{{\rm cr}}<S_{0}$, as the soliton cannot have an amplitude
larger than $\sqrt{S_{{\rm cr}}}$, the amplitude takes this critical value.
The soliton is singular in this case, being a {\it cuspon} (see details
below), but, nevertheless, it is an absolutely relevant solution. The
remaining possibilities are that either $S_{{\rm cr}}<0$ and $S_{0}>0$, or
vice versa; then the soliton may only be, respectively, regular or singular.
Of course no soliton exists if both $S_{0}$ and $S_{{\rm cr}}$ are negative.
Further, using the symmetries of the equations, it is readily shown that for
all these soliton solutions, $S(x)$ is symmetric about its center, which may
be set at $x=0$, that is, $S(x)$ is an even function of $x$. For the cuspon
solutions, and for those regular solutions whose squared amplitude is $
S_{01} $, it can also be shown that the phase variable $\psi (x)=\phi
(x)-\pi $ and $U_{3}(x)$ are odd functions of $x$, while for those regular
solutions whose squared amplitude is $S_{02}$ the phase variable $\phi (x)$
and $U_{3}(x)$ are, respectively, odd and even functions of $x$.

It is now necessary to determine which parameter combinations in the set 
($\omega ,\omega _{0},\alpha $) permit the options described above. The most
interesting case occurs when $\omega _{0}>\omega $ (so that $\omega $
belongs to the lower gap, see Fig. 2) and $\alpha >0$ (the latter condition
always holds in the applications to nonlinear optics). In this case, it can
be shown that the root $S_{02}$ of Eq. (\ref{quadr}) is not relevant, and
the options are determined by the competition between $S_{01}$ and $S_{{\rm
cr}}$. The soliton is a cuspon ($0<S_{{\rm cr}}<S_{01}$) if 
\begin{equation}
\alpha (\omega _{0}-\omega )\,<\,4(1-\omega ).  \label{criterion}
\end{equation}
In effect, the condition (\ref{criterion}) sets an upper bound on $\alpha$
for given $\omega _{0}$ and $\omega $. In particular, this condition is
always satisfied if $0<\alpha <4$.

If, on the other hand, the condition (\ref{criterion}) does not hold (i.e., 
$0<S_{01}<S_{{\rm cr}}$), we obtain a regular soliton. In a less physically
relevant case, when again $\omega _{0}>\omega $ but $\alpha <0$, cuspons do
not occur (as this time $S_{{\rm cr}}<0$, see Eq. (\ref{critical}), and only
regular solitons may exist.

Next we proceed to the case $\omega _{0}<\omega $, so that $\omega $ is
located in the upper gap of the linear spectrum. For $\alpha >0$, we have 
$S_{{\rm cr}}<0$, hence only regular solitons may occur, and indeed in this
case there is always at least one positive root $S_{0}$, so a regular
soliton does exist. If $\alpha <0$, then we have $S_{{\rm cr}}>0$, but if 
$\omega _{0}<1-\kappa ^{2}$ (when also $\omega <1$), there is at least one
positive root $S_{0}<S_{{\rm cr}}$; thus, only a regular soliton can exist
in this case too. On the other hand, if $\alpha <0$ and $\omega
_{0}>1-\kappa ^{2}$ (and then $\omega >1$), there are no positive roots 
$S_{0}$, and so only cuspons occur.

Let us now turn to a detailed description of the cuspon's local structure
near its center, when $S$ is close to $S_{{\rm cr}}$. From the above
analysis, one sees that cuspons occur whenever $\omega $ lies in the lower
gap, with $\omega _{0}>\omega $ and $\alpha >0$, so that the criterion (\ref
{criterion}) is satisfied, or when $\omega $ lies in the upper gap with $
1-\kappa ^{2}<\omega _{0}<\omega $ and $\alpha <0$. To analyze the structure
of the cuspon, we first note that, as it follows from Eq. (\ref{integral}),
one has $\cos \phi =-1$ (i.e., $\phi =\pi $) when $S=S_{{\rm cr}}$, which
suggest to set 
\begin{equation}
S_{{\rm cr}}-S\,\equiv \,\delta \cdot \,\kappa ^{2}R,\qquad 1+\cos \phi
\,\equiv \,\delta \,\cdot \rho ,  \label{stretch}
\end{equation}
where $\delta $ is a small positive parameter, and the stretched variables 
$R $ and $\rho $ are positive. At the leading order in $\delta $, it then
follows from Eq. (\ref{integral}) that $\rho \,=\rho _{0}R$, where 
\begin{equation}
\rho _{0}\,\equiv \,\alpha ^{3}(S_{01}-S_{{\rm cr}}).  \label{rho}
\end{equation}
As it follows from the above analysis, $\rho _{0}$ is always positive for a
cuspon. We also stretch the spatial coordinate, defining $x\,\equiv \delta
^{3/2}\kappa ^{2}y$, the soliton center being at $x=0$. Since $S(x)$ is an
even function of $x$, it is sufficient to set $x>0$ in this analysis. Then,
on substituting the first relation from Eq. (\ref{stretch}) into Eq. (\ref
{S1}), we get, to the leading order in $\delta $, an equation 
\begin{equation}
R\left( dR/dy\right) ^{2}\,=\rho _{0}S_{{\rm cr}}^{2}/\alpha ^{2}\equiv
\,K^{2},  \label{R}
\end{equation}
so that 
\begin{equation}
R\,=\,\left( 3Ky/2\right) ^{2/3}.  \label{stretch3}
\end{equation}
Note that in the original unstretched variables, the relation (\ref{stretch3}
) shows that, near the cusp, 
\begin{equation}
S_{{\rm cr}}-S(x)\approx (3K\kappa x/2)^{2/3},  \label{cusp}
\end{equation}
\begin{equation}
dS/dx\approx (2/3)^{1/3}\left( K\kappa \right) ^{2/3}\cdot x^{-1/3},
\label{singular}
\end{equation}
and it follows from Eq. (\ref{explicit}) that $U_{3}$ is unbounded near the
cusp, 
\begin{equation}
U_{3}\,\approx \,(S_{{\rm cr}}/\alpha )(2\alpha \rho _{0}K^{2}/3\kappa
x)^{1/3}.  \label{U3singular}
\end{equation}
In particular, Eq. (\ref{singular}) implies that, as $K\kappa $ decreases,
the cusp gets localized in a narrow region where $|x|\,\,_{\sim
}^{<}\,K^{2}\kappa ^{2}$ (outside this region, $\left| dS/dx\right| $ is
bounded and shows no cusp). Note that this limit can be obtained either as 
$\kappa ^{2}\rightarrow 0$, or as $\rho _{0}\rightarrow 0$ (recall that $\rho
_{0}$ is defined in Eq. (\ref{rho})).

An example of the cuspon is shown in Fig. 3. Although the first derivative
in the cuspon is singular at its center, as it follows from Eq. (\ref
{singular}) [see also Fig. 3(a)], and its $U_{3}$ component diverges at $
x\rightarrow 0$ as per Eq. (\ref{U3singular}), it is easily verified that
the value of the Hamiltonian (\ref{H}) [and, obviously, the value of the
norm (\ref{E}) too] is finite for the cuspon solution. These solitons are
similar to cuspons found as exact solutions to the Camassa-Holm (CH)
equation \cite{CH,cuspon}, which have a singularity of the type $|x|^{1/3}$
or $|x|^{2/3}$ as $|x|\rightarrow 0$, cf. Eqs. (\ref{cusp}) and (\ref
{singular}). The CH equation is integrable, and it is degenerate in the
sense that it has no linear terms except for $\partial u/\partial t$ (which
makes the existence of the solution with a cusp singularity possible). Our
three-wave system (\ref{1}) - (\ref{3}) is not degenerate in that sense;
nevertheless, the cuspon solitons are possible in it because of the model's
multicomponent structure: the elimination of the third component generates
the non-polynomial nonlinearity in Eqs. (\ref{U1}), (\ref{U2}), and,
finally, in Eqs. (\ref{S}) and (\ref{S1}), which gives rise to the cusp. It
is noteworthy that, as well as the CH model, ours gives rise to two
different {\it coexisting} families of solitons, viz., regular ones and
cuspons. It will be shown below that the solitons of both types may be
stable.

Of course, the presence of the singularities in $U_{3}(x)$ and $dS/dx$ at 
$x\rightarrow 0$ suggests that higher-order terms, such as the higher-order
dispersion, should be taken into regard in this case. The fact that the
cuspon's Hamiltonian converges despite these singularities, as well as a
direct analysis, suggest that such higher-order terms will smooth the shape
of the cuspon in a very narrow layer for small $x$, allowing for large but
not diverging values of the fields. However, the small higher-order terms
will not essentially alter the global shape of the cuspons. In the next
section we will show that, in fact, the genuine generic singular solitons
are (in the presence of the SPM terms) peakons, for which the singularities
are much weaker, hence the latter issue is still less significant. Besides
that, it appears to be an issue of principal interest to understand what
types of solitons the system may generate without intrinsic dispersion (cf.
the situation for the traditional GMTM, in which the spectrum of soliton
solutions is completely altered by the addition of intrinsic dispersion 
\cite{Alan}).

In the special case $\kappa \ll 1$, when the third component is weakly
coupled to the first two ones in the linear approximation (in terms of the
optical model represented by Fig. 1, it is the case when the subgratings
shown by the dashed lines are very weak), straightforward inspection of the
above results shows that the cuspons look like {\it peakons}; that is,
except for the above-mentioned narrow region of the width $|x|\sim \kappa
^{2}$, where the cusp is located, they have the shape of a soliton with a
discontinuity in the first derivative of $S(x)$ and a jump in the phase $
\phi (x)$, which are the defining features of peakons (\cite{CH,peakon}). A
principal difference of true peakons from cuspons is that the first
derivative does not diverge inside a peakon, but is of course, discontinuous.

An important result of our analysis is that the family of solitons obtained
in the limit $\kappa \rightarrow 0$ is drastically different from that in
the model where one sets $\kappa =0$ from the very beginning. In particular,
in the most relevant case, with $\omega _{0}>\omega $ and $\alpha >0$, the
family corresponding to $\kappa \rightarrow 0$ contains regular solitons
whose amplitude is smaller than $\sqrt{S_{{\rm cr}}}$; however, the solitons
whose amplitude at $\kappa =0$ is larger than$\sqrt{S_{{\rm cr}}}$, i.e.,
the ones whose frequencies belong to the range (\ref{criterion}) [note that
the definition of $S_{{\rm cr}}$ does not depend on $\kappa $ at all, see
Eq. (\ref{critical})], are replaced by the peakons which are constructed in
a very simple way: drop the part of the usual soliton above the critical
level $S=S_{{\rm cr}}$, and bring together the two symmetric parts which
remain below the critical level, see Fig. 3(b).

It is interesting that peakons are known as exact solutions to a version of
the integrable CH equation slightly different from that which gives rise to
the cuspons. As well as in the present system, in that equation the peakons
coexist with regular solitons \cite{peakon}. In the next subsection, we
demonstrate that the peakons, which are found only as limit-form solutions
in the no-SPM case $\sigma _{3}=0$, become generic solutions in the case $
\sigma _{3}\neq 0$.

\subsection{Peakons, the case $\protect\sigma_3 \ne 0$}

A natural question is whether the cuspon solutions are {\em structurally
stable}, i.e., if they will persist on inclusion of terms that were absent
in the analysis presented above (the other type of the stability, viz.,
dynamical stability against small initial perturbations, will be considered
in the next section). Here, we address this issue by restoring the SPM term
in Eq. (\ref{3}), that is, we now set $\sigma _{3}\neq 0$, but assume that
it is a small parameter. Note that, in the application to nonlinear optics,
one should expect that $\sigma _{3}>0$, but there is no such a restriction
on the sign of $\sigma _{3}$ in the application to the flow of a
density-stratified fluid. We still keep $\sigma _{1}=0$, as the inclusion of
the corresponding SPM terms in Eqs. (\ref{1}) and (\ref{2}) amounts to
trivial changes both in the above analysis, and in that presented below. On
the other hand, we show below that the inclusion of the SPM term in Eq. (\ref
{3}) is a structural perturbation which drastically changes the character of
the soliton solutions.

In view of the above results concerning the cuspons, we restrict our
discussion here to the most interesting case when $S(x)$ is an even function
of $x$, while $\psi (x)=\phi (x)-\pi $ and $U_{3}(x)$ are odd functions. In
principle, one can use the relations (\ref{U3real}) and (\ref{integral}) to
eliminate $\phi $ and $U_{3}$ and so obtain a single equation for $S$ (a
counterpart to Eq. (\ref{S1})), as it was done above when $\sigma _{3}=0$.
However, when $\sigma _{3}\neq 0$, it is not possible to do this explicitly.
Instead, we shall develop an asymptotic analysis valid for $x\rightarrow 0$,
which will be combined with results obtained by direct numerical integration
of Eqs. (\ref{phi}) and (\ref{S}), subject of course to the constraints (\ref
{U3real}) and (\ref{integral}). Since singularities only arise at the center
of the soliton (i.e., at $x=0$) when $\sigma _{3}=0$, it is clear that the
introduction of a small $\sigma _{3}\neq 0$ will produce only a small
deformation of the soliton solution in the region where $x$ is bounded away
from zero.

First, we consider regular solitons. Because the left-hand side of Eq. (\ref
{U3real}) is not singular at any $x$, including the point $x=0$, when $
\sigma _{3}=0$, we expect that regular solitons survive a perturbation
induced by $\sigma _{3}\neq 0$. Indeed, if there exists a regular soliton,
with $S_{0}\equiv S(x=0)$, and $\phi (x=0)=\pi $ and $U_{3}(x=0)=0$, it
follows from Eq. (\ref{integral}) that the soliton's amplitude remains
exactly the same as it was for $\sigma _{3}=0$, due to the fact that the
regular soliton has $U_{3}(x=0)=0$.

Next, we turn to the possibility of singular solutions, that is, cuspons or
peakons. Since we are assuming that $S_{0}=S(x=0)$ is finite, and that $\phi
(x=0)=\pi $, it immediately follows from Eq. (\ref{U3real}) that when $
\sigma _{3}\neq 0$, $U_{3}$ must remain finite for all $x$, taking some
value $U_{0}\neq 0$, say, as $x\rightarrow +0$. Since $U_{3}$ is an odd
function of $x$, and $U_{0}\neq 0$, there must be a discontinuity in $U_{3}$
at $x=0$, i.e., a jump from $U_{0}$ to $-U_{0}$. This feature is in marked
contrast to the cuspons for which $U_{3}$ is infinite at \ the center, see
Eq. (\ref{U3singular}). Further, it then follows from Eq. (\ref{S}) that, as 
$x\rightarrow 0$, there is also a discontinuity in $dS/dx$, with a jump from 
$2\kappa U_{0}\sqrt{S_{0}}$ to $-2\kappa U_{0}\sqrt{S_{0}}$. Hence, if we
can find soliton solutions of this type, with $U_{0}\neq 0$, they are
necessarily {\it peakons}, and we infer that cuspons do {\em not} survive
the structural perturbation induced by $\sigma _{3}\neq 0$.

Further, if we assume that $U_{0}\neq 0$, then Eq. (\ref{U3real}), taken in
the limit $x\rightarrow 0$, immediately shows that 
\begin{equation}
2\alpha (S_{{\rm cr}}-S_{0})\,=\,\sigma _{3}U_{0}^{2}  \label{U0}
\end{equation}
(recall that $S_{{\rm cr}}$ is defined by Eq. (\ref{critical})). Next, the
Hamiltonian relation (\ref{integral}), also taken in the limit $x\rightarrow
0$, shows that 
\begin{equation}
-\frac{\rho _{0}}{\alpha }S_{0}-\alpha ^{2}S_{0}(S_{{\rm cr}}-S_{0})\,=\,
\frac{1}{2}\sigma _{3}U_{0}^{4},  \label{H0}
\end{equation}
where we have used Eq. (\ref{U0}) [recall that $\rho _{0}$ is defined by Eq.
(\ref{rho})]. Elimination of $U_{0}$ from Eqs. (\ref{U0}) and (\ref{H0})
yields a quadratic equation for $S_{0}$, whose positive roots represent the
possible values of the peakon's amplitude.

We recall that for a cuspon which exists at $\sigma _{3}=0$ one has $\rho
_{0}>0$, i.e., the amplitude of the corresponding formal regular soliton
exceeds the critical value of the amplitude, see Eq. (\ref{rho}). Then, if
we retain the condition $\rho _{0}>0$, it immediately follows from Eqs. (\ref
{U0}) and (\ref{H0}) that no peakons may exist if the SPM coefficient in Eq.
(\ref{3}) is positive, $\sigma _{3}>0$. Indeed, Eq. (\ref{U0}) shows that 
$S_{{\rm cr}}-S_{0}>0$ if $\sigma _{3}>0$, which, along with $\rho _{0}>0$,
leads to a contradiction in the relation (\ref{H0}).

Further, it is easy to see that a general condition for the existence of
peakons following from Eqs. (\ref{U0}) and (\ref{H0}) is 
\begin{equation}
\sigma _{3}\rho _{0}<0\,,  \label{necessary_for_peakon}
\end{equation}
hence peakons are possible if $\sigma _{3}<0$, or if we keep $\sigma _{3}>0$
but allow $\rho _{0}<0$. In the remainder of this section, we will show that
peakons may exist only if $\rho _{0}>0$. Hence, it follows from the
necessary condition (\ref{necessary_for_peakon}) that peakons may indeed be
possible solely in the case $\sigma _{3}<0$. On the other hand, regular
solitons do exist in the case $\sigma _{3}>0$ (i.e., in particular, in
nonlinear-optics models), as they have $U_{0}=0$, hence neither Eq. (\ref{U0}
) nor its consequence in the form of the inequality (\ref
{necessary_for_peakon}) apply to regular solitons. The existence of (stable)
peakons for $\sigma _{3}<0$, and of (also stable) regular solitons for $
\sigma _{3}>0$ will be confirmed by direct numerical results presented in
the next section.

To obtain a necessary condition (which will take the form of $\rho _{0}>0$)
for the existence of the peakons, we notice that the existence of any
solitary wave implies the presence of closed dynamical trajectories in the
phase plane of the corresponding dynamical system, which is based on the
ordinary differential equations (\ref{phi}) and (\ref{S}), supplemented by
the constraint (\ref{U3real}). Further, at least one stable fixed point (FP)
must exist inside such closed trajectories, therefore the existence of such
a stable FP is a necessary condition for the existence of any solitary wave.

The FPs are found by equating to zero the right-hand sides of Eq. (\ref{phi}
) and (\ref{S}), which together with Eq. (\ref{U3real}) give three equations
for the three coordinates $\phi ,S$ and $U_{3}$ of the FP. First of all, one
can find a trivial unstable FP of the dynamical system, 
\[
\cos \phi =-\frac{\omega +\kappa ^{2}/(\omega _{0}-\omega )}{1+\kappa
^{2}/(\omega _{0}-\omega )}\,,\quad S\,=\,0\,, 
\]
which does not depend on $\sigma _{3}$. Then, three nontrivial FPs can be
found, with their coordinates $\phi _{*}$, $S_{*}$ and $U_{3*}$
given by the following expressions: 
\begin{equation}
\phi _{\ast }^{(1)}\,=\,\pi ,\quad S_{\ast }^{(1)}\,=\frac{1-\omega }{
\alpha^{2}}\,=\,\frac{1}{2}S_{01},\quad U_{3\ast }^{(1)}\,=\,0,  \label{FP1}
\end{equation}
\begin{equation}
\phi _{\ast }^{(2)}\,=\,\pi ,\quad (2-\sigma _{3})S_{\ast }^{(2)}\,=
2S_{{\rm cr}}-\frac{\sigma _{3}}{2}S_{01},\quad 
(2-\sigma _{3})\left[ \alpha
U_{3\ast}^{(2)}\right] ^{2}\,=\rho _{0}-\alpha ^{3}S_{{\rm cr}}\,,
\label{FP2}
\end{equation}
\[
(2-\sigma _{3})S_{*}^{(3)}\,=2S_{{\rm cr}}-\frac{1}{2}\sigma _{3}S_{01}+ 
\frac{\kappa ^{2}}{\alpha },\quad (2-\sigma _{3})\left[ \alpha U_{3\ast
}^{(3)}\right] ^{2}\,=\rho _{0}-\alpha ^{3}S_{{\rm cr}}-\alpha ^{2}\kappa
^{2}\,,\newline
\]
\begin{equation}
\cos \left( \phi _{\ast }^{(3)}/2\right) =-\frac{1}{2}\kappa
U_{3\ast}^{(3)}\,/\sqrt{S_{\ast }^{(3)}}\,\,,  \label{FP3}
\end{equation}
where the superscript is a number label for the FP. To be specific, we now
consider the case of most interest, when both $S_{01}>0$ and $S_{{\rm cr}}>0$
. In this case, the FP given by Eqs. (\ref{FP1} ) exists for all $\sigma
_{3} $ and all $\rho _{0}$. However, for small $\sigma _{3}$ (in fact $
\sigma _{3}<2$ is enough) and small $\kappa $, the FPs given by Eqs. (\ref
{FP2}) and (\ref{FP3}) exist only when $\rho _{0}>0$. Indeed, they exist
only for $\rho _{0}>\alpha ^{3}S_{01}$ and $\rho _{0}>\alpha
^{3}S_{01}+\kappa ^{2}$, respectively, or, on using the definition (\ref{rho}
) of $\rho _{0}$, when $S_{01}>2S_{{\rm cr}}$ and $S_{01}>2S_{{\rm cr}
}+\kappa ^{2}/\alpha $, respectively.

Let us first suppose that $\rho _{0}<0$. Then there is only the single
non-trivial FP, namely the one given by Eqs. (\ref{FP1}). This FP is clearly
associated with the regular solitons, whose amplitude at the crest is $
S_{01} $. Hence, we infer that for $\rho _{0}<0$ there are no other
solitary-wave solutions, and in particular, no peakons (and no cuspons when $
\sigma _{3}=0$ either, in accordance with what we have already found in
subsection 2.3 above). When combined with the necessary condition (\ref
{necessary_for_peakon}) for the existence of peakons, we infer that there
are no peakons when $\sigma _{3}>0$, thus excluding peakons from
applications to the nonlinear-optics models, where this SPM coefficient is
positive. However, peakons may occur in density-stratified fluid flows,
where there is no inherent restriction on the sign of $\sigma _{3}$. This
case is considered below, but first we note that in the case $\rho _{0}<0$
and $\sigma _{3}>0$ (which includes the applications to nonlinear optics),
the same arguments suggest that there may be {\it periodic} solutions with a
peakon-type discontinuity at the crests; indeed, our numerical solutions of
the system (\ref{phi},\ref{S}) show that this is the case.

Next, we suppose that $\rho _{0}>0$. First, if $S_{01}<2S_{{\rm cr}}$ then
there is again the single non-trivial FP given by (\ref{FP1}). But now, by
analogy with the existence of cuspons when $\rho _{0}>0$ and $\sigma _{3}=0$
, , we infer that the solitary wave solution which is associated with this
fixed point is a peakon, whose squared amplitude $S_{0}$ for small $\sigma
_{3}$ is close to $S_{{\rm cr}}$, while the FP has $S_{\ast
}^{(1)}=S_{01}/2<S_{{\rm cr}}$.

If, on the other hand, $S_{01}>2S_{{\rm cr}}$, the FPs given by Eqs.(\ref
{FP2}) and (\ref{FP3}) become available as well. We now infer that the
peakon solitary-wave solution continues to exist, and for sufficiently small 
$\sigma _{3}$ and $\kappa $ it is associated with the FP given by Eq. (\ref
{FP2}). Although Eq. (\ref{FP2}) implies that $S_{\ast }^{(2)}\approx S_{
{\rm cr}}$, and the peakon's squared amplitude $S_{0}$, determined by Eqs. (
\ref{U0}) and (\ref{H0}), is also approximately equal to $S_{{\rm cr}}$, we
nevertheless have $S_{0}>S_{\ast }^{(2)}$ as required. Note that, in the
present case, the FPs given by Eqs. (\ref{FP1}) and (\ref{FP3}) lie outside
the peakon's homoclinic orbit. In Fig. 4, we show a plot of a typical peakon
obtained, in this case, by numerical solution Eqs. (\ref{phi}) and (\ref{S}).

\section{Numerical results}

\subsection{Simulation techniques}

The objectives of direct numerical simulations of the underlying equations 
(\ref{1}) - (\ref{3}) were to check the dynamical stability of regular
solitons, cuspons, and peakons in the case $\sigma _{3}=0$, and the
existence and stability of peakons in the more general case, $\sigma
_{3}\neq 0$. Both finite-difference and pseudo-spectral numerical methods
have been used, in order to check that the same results are obtained by
methods of both types. We used semi-implicit Crank-Nicholson schemes, in
which the nonlinear terms were treated by means of the Adams-Bashforth
method.

The presence of singularities required a careful treatment of cuspon and
peakon solutions. To avoid numerical instabilities due to discontinuities,
we found it, sometimes, beneficial to add an artificial weak high-wavenumber
viscosity to the pseudospectral code. This was done by adding linear damping
terms to the right-hand side of Eqs. (\ref{1}), (\ref{2}) and (\ref{3}),
which have the form $-i\nu (k)k^{2}\hat{u_{n}}$ in the Fourier
representation, where $\hat{u_{n}}$ is the Fourier transform of $u_{n}$ 
($n=1,2,3$). The high-pass filter viscosity $\nu (k)$ suppresses only high
wavenumbers and does not act on others. In particular, we chose 
\[
\nu (k)=\left\{ 
\begin{array}{ll}
0\,, & {\rm if}\,\,|k|<\frac{5}{16}K \\ 
\eta (\frac{16|k|}{K}-5), & {\rm if}\,\,\frac{5}{16}K<|k|<\frac{3}{8}K \\ 
\eta ,\, & {\rm if}\,\,|k|>\frac{3}{8}K
\end{array}
\right. \;, 
\]
where $K$ is the largest wavenumber in the actual numerical scheme, and 
$\eta $ is a small viscosity coefficient. We have found that $\eta \sim
10^{-5}$ was sufficient to avoid Gibbs' phenomenon in long-time simulations.

When instabilities occur at a singular point (cusp or peak), it is hard to
determine whether the instability is a real one, or a numerical artifact.
Therefore, we checked the results by means of a finite-difference code which
used an adaptive staggered grid; motivated by the analysis of the vicinity
of the point $x=0$ reported above, we introduced the variable $\xi \equiv
x^{2/3}$ to define an adaptive grid, and also redefined $U_{3}\equiv \sqrt{
\xi}\widetilde{U}_{3}$. In these variables, the cusp becomes a regular
point. This approach was solely used to check the possible occurrence of
numerical instabilities.

In the following subsections we present typical examples of the numerical
results for both cases considered above, viz., $\sigma _{3}=0$ and $\sigma
_{3}<0$, when, respectively, cuspons and peakons are expected.

\subsection{The case $\protect\sigma _{3}=0$}

First, we report results obtained for the stability of regular solitary
waves in the case $\sigma _{3}=0$. As initial configurations, we used the
corresponding stationary solutions to Eqs. (\ref{phi}) and (\ref{S}). To
test the stability of the regular solitary waves, we added small
perturbations to them. As could be anticipated, the regular solitary wave
sheds off a small dispersive wave train and relaxes to a stationary soliton,
see Fig. 5 (for a more detailed illustration of the generation of small
radiated waves by a soliton, see also Fig. 8 below). If, however, a regular
soliton is taken as an initial condition for parameter values inside, but
close to the border of the cuspon region, it does not become unstable in
this slightly modified section of parameter space (which only supports
cuspons), but instead this soliton exhibits persistent internal vibrations,
see an example in Fig. 6. These and many other simulations clearly show that
the regular soliton is always stable, and, close to the parameter border
with cuspons, it has a persistent internal mode.

It was shown analytically above that Eqs. (\ref{U1}) and (\ref{U2}) ( with 
$\sigma _{3}=0$) support peakons when 
$\rho _{0}>0$ and $\rho _{0}\kappa ^{2}$
is very small. Direct simulations show that peakons do exist in this case,
and they may be either unstable or stable. In the case when they are
unstable, a high-wavenumber instability develops around the central peak. In
Fig. 7, we display the time evolution of a typical stable peakon.

Next, we look at what happens if we take a regular soliton as an initial
condition in a section of the parameter space which supports only stable
peakons. This enables us to study the competition of the structural
stability of regular solitons (as confirmed in Fig. 6) and the stability of
peakons. The initial condition is taken as a stationary regular soliton in
the parameter region (close to the boundary of the peakon region) with $
\rho_0 < 0$, whereas the simulations are run for values of the parameters
corresponding to $\rho_0>0$, which only admits peakons and excludes regular
solitons. Unlike the case shown in Fig. 6, the time evolution now does not
exhibit internal vibrations. Instead, the pulse slowly decays into
radiation. This outcome can be explained by the fact that the peakon's norm
(see Eq.(\ref{E})) turns out to be larger than that of the initial pulse in
this case, hence its rearrangement into a stable peakon is not possible. 
An essential result revealed by the simulations is that cuspons may also be 
{\em stable}, a typical example being displayed in Fig. 8. In this figure,
one can see a small shock wave which is initially generated at the cuspon's
crest. It seems plausible that this shock wave is generated by some initial
perturbation which could be a result of the finite mesh size in the
finite-difference numerical scheme employed for the simulations. In fact,
the emission of a small-amplitude shock wave is quite a typical way of the
relaxation of both cuspons and peakons to a final stable state. To make it
sure that the shock wave is not an artifact generated by the numerical
scheme, we have checked that its shape does not change with the increase of
the numerical accuracy.

To further test the stability of the cuspons and peakons, in many cases we
allowed the initially generated shock wave to re-enter the integration
domain (due to periodic boundary conditions used in the simulations) and
interact again with the cuspon or peakon. As a result, the stability of the
solitons of these types has been additionally confirmed. An example of the
spatial profile of the cuspon established after a long evolution is shown in
Fig. 9. Both the stability of the cuspon, and the presence of a tiny shock
wave are evident in the figure.

However, unlike the regular solitons, which were found to be always stable,
the cuspons are sometimes unstable. Typically, their instability triggers
the onset of spatiotemporal collapse, i.e., formation of a singularity in a
finite time (see a discussion of the feasible collapse in systems of the
present type, given in the Introduction). Simulations of the collapse were
possible with the use of an adaptive grid. A typical example of the collapse
is shown in Fig. 10, where the inset shows that (within the numerical
accuracy available) the amplitude of the collapsing pulse indeed diverges in
a finite time.

However, the collapse is not the only possible outcome of the instability.
In some other cases, which are not displayed here, the instability of
peakons could be quite weak, giving rise to their rearrangement into regular
solitons by the shedding of a small amount of radiation.

\subsection{The case $\protect\sigma _{3} \neq 0$}

The predictions of the analysis developed above for the most general case,
when the SPM terms are present in the model ($\sigma _{3}\neq 0$), were also
checked against direct simulations. As a result, we have found, in accord
with the predictions, that only regular solitons exist in the case $\sigma
_{3}>0$, while in the case $\sigma _{3}<0$ both regular solitons and peakons
have been found as generic solutions. Further simulations, details of which
are not shown here, demonstrate that both regular solitons {\em and} peakons
are stable in this case.

\section{Conclusion}

In this work, we have introduced a generic model of three waves coupled by
linear and nonlinear terms, which describes a situation when three
dispersion curves are close to an intersection at one point. The model was
cast into the form of a system of two waves with opposite group velocities
that, by itself, gives rise to the usual gap solitons, which is further
coupled to a third wave with zero group velocity (in the laboratory
reference frame). Situations of this type are quite generic, being
realizable in various models of nonlinear optics, density-stratified fluid
flows, and in other physical contexts. In particular, two versions (temporal
and spatial) of a nonlinear-optical model, which is based on a waveguide
carrying the triple spatial Bragg grating, have been elaborated in the
Introduction. Our consideration was focussed on zero-velocity solitons. In a
special case when the self-phase modulation (SPM) is absent in the equation
for the third wave, soliton solutions were found in an exact form. It was
shown that there are two coexisting generic families of solitons: regular
solitons and cuspons. In the special case when the coefficient of the linear
coupling between the first two waves and the third one vanishes, cuspons are
replaced by peakons. Direct simulations have demonstrated that the regular
solitons are stable (in the case when the regular soliton is close to the
border of the cuspon region, it has a persistent internal mode). The cuspons
and peakons may be both stable and unstable. If they are unstable, they
either shed off some radiation and rearrange themselves into regular
solitons, or, in most typical cases, the development of the cuspon's
instability initiates onset of spatiotemporal collapse. Actually, the
present system gives the first explicit example of collapse in
one-dimensional gap-soliton models.

The most general version of the model, which includes the SPM term in the
equation for the third wave, has also been considered. Analysis shows that
cuspons cannot exist in this case, i.e., cuspons, although being possibly
dynamically stable, are structurally unstable. However, depending on the
signs of the SPM coefficient, and some combination of the system's
parameters, it was shown that a generic family of peakon solutions may exist
instead. In accord with this prediction, the peakons have been found in
direct simulations. The peakons, as well as the regular solitons, are stable
in the system including the SPM term. We stress that peakons are physical
solutions, as they have all their field components and their first
derivatives finite.

The next step in the study of this system should be consideration of moving
solitons, which is suggested by the well-known fact that the usual two-wave
model gives rise to moving gap solitons too \cite{review}. However, in
contrast to the two-wave system, one may expect a drastic difference between
the zero-velocity and moving solitons in the present three-wave model. This
is due to the reappearance of a derivative term in Eq. (\ref{3}), when it is
written for a moving soliton, hence solitons which assume a singularity or
jump in the $U_{3}$ component, i.e., both cuspons and peakons, cannot then
exist. Nevertheless, one may expect that slowly moving solitons will have
approximately the same form as the cuspons and peakons, with the singularity
at the central point replaced by a narrow transient layer with a large
gradient of the $U_{3}$ field. Detailed analysis of the moving solitons is,
however, beyond the scope of this work.

\section*{Acknowledgements}

We would like to thank Tom Bridges, Gianne Derks and Sebastian Reich for
valuable discussions. B.A.M. and G.A.G. appreciate hospitality of the
University of Loughborough (UK). The work of G.A.G. is supported by a
European Commission Grant, contract number HPRN-CT-2000-00113, for the
Research Training Network {\it Mechanics and Symmetry in Europe\/} (MASIE)

\newpage

Fig. 1. A schematic representation of the optical model that gives rise to
the three linearly coupled waves in a $\chi ^{(3)}$ waveguide. The figure
shows either the planar waveguide with the triple Bragg grating, in the case
of the temporal evolution of the fields, or the transverse cross section of
the bulk medium of the layered (photonic-crystal-fiber) type, in the case of
the spatial evolution along the coordinate $z$ perpendicular to the plane of
the figure. The triangles formed by bold arrows illustrate how the linear
couplings between the three waves, whose Poynting vectors are represented by
the arrows, are induced by the Bragg reflections on the three gratings. The
difference between the gratings represented by the continuous and dashed
lines is in their strength (refractive-index contrast).

Fig. 2. Dispersion curves produced by Eq. (\ref{dispersion}) in the case 
$\kappa =0.5$: (a) $\omega _{0}<1-\kappa ^{2}$; (b) $\omega _{0}>1$. The
dashed line in each panel is $\omega =\omega _{0}$. The case with $1-\kappa
^{2}<\omega _{0}<1$ is similar to the case (a) but with the points $\omega
_{+}$ and $1$ at $k=0$ interchanged.

Fig. 3. The shape of the cuspon for $\alpha =2.0$, $\omega _{0}=0.1$, 
$\omega =-0.5$, and (a) $\kappa =0.5$, i.e., in the general case, and (b) 
$\kappa =0.1$, i.e., for small $\kappa$. In the case (b) we also show the
usual gap soliton (by the dashed line), the part of which above the critical
value $S=S_{{\rm cr}}$ (shown by the dotted line) should be removed and the
remaining parts brought together to form the peakon corresponding to 
$\rho_0\kappa^2 \rightarrow 0$.

Fig. 4. The shape of the peakon for the case when $\sigma _{3}<0$, where we
plot $|U_{1}|^{2}$. The parameters are $\sigma _{3}=-0.01$, $\kappa =0.1$, 
$\alpha =2.0$, $\omega _{0}=0.1$, and $\omega =-0.5$. In this case, $\rho
_{0}=4.8$.

Fig. 5. The shape of an initially perturbed regular soliton in the case 
$\sigma _{3}=0$ at $t=5$, which illustrates the stabilization of the soliton
through the shedding of small-amplitude radiated waves. The plot displayed
is the field ${\rm Re}U_{1}(x)$. The parameters are $\kappa =0.01,\alpha
=1.0,\omega _{0}=0.2$, and $\omega =0.9$.

Fig. 6. Internal vibrations of an initially-perturbed regular soliton, which
was taken close to the parameter boundary of the cuspon region. The plot
shows the squared amplitude $a\equiv \left| U_{1}(x=0)\right| ^{2}$ of the 
$U_{1}(x)$ field versus time. The parameters are $\kappa =0.01,\alpha
=1.9,\omega _{0}=1.5$, and $\omega =0.5$, with $\rho _{0}=0.095$ (see Eq. 
(\ref{rho}).

Fig. 7. An example of a stable peakon. The plot shows the field $\left|
U_{1}\right| ^{2}$ versus $x$ and $t$. The parameters are $\kappa
=1.0,\alpha =1.95,\omega _{0}=1.5$, and $\omega =0.5$, with $\rho
_{0}=0.04875$.

Fig. 8. An example of a stable cuspon. The plot shows the field $\left|
U_{1}\right| ^{2}$ versus $x$ and $t$. The parameters are $\kappa
=1.0,\alpha =1.0,\omega _{0}=1.5$, and $\omega =0.5$, with $\rho _{0}=0.5$.

Fig. 9. The spatial profile of the stable cuspon at $t=20$. The parameters
are the same as in Fig. 8.

Fig. 10. The spatial profile is shown for an unstable cuspon in 
terms of ${\rm Im}U_{1}$ at $t=10^{-3}$. The inset depicts the 
time-evolution of the maximum value of $\left|
U_{1}\right|^{2}$. The transition to collapse is clearly seen as 
an explosive temporal behavior of the amplitude.
The parameters are $\kappa =0.01,\alpha =1.1,\omega _{0}=0.1$, and
$\omega =-0.3$, with $\rho _{0}=2.618$.

\end{document}